\documentclass[article]{aastex631}

\usepackage{mathrsfs}
\usepackage{amsmath}

\begin{document}

\title{The Correspondence Between Leaky-Box and Diffusion Models of Cosmic-Ray Propagation}

\correspondingauthor{Ramanath Cowsik}
\email{cowsik@wustl.edu}

\author{Ramanath Cowsik}
\affiliation{Washington University in Saint Louis Department of Physics and the McDonnell Center for the Space Sciences \\
1 Brookings Drive, Saint Louis, MO 63130, USA}

\author[0000-0002-4987-0678]{Dawson Huth}
\affiliation{Washington University in Saint Louis Department of Physics and the McDonnell Center for the Space Sciences \\
1 Brookings Drive, Saint Louis, MO 63130, USA}

\begin{abstract}

The leaky-box model and the attendant concept of path-length distribution of cosmic rays were invented in the mid-1960's. Even though versatile computational packages such as GALPROP and DRAGON with the diffusion approach are now available for analyzing cosmic ray data, the concepts of the leaky-box and path-length distribution continue to be adopted extensively. We show here mathematically that there is a close correspondence between the two approaches: The path-length or resident-time of the leaky-box models are similar to 'impulse response functions' of complex dynamical systems and are intuitively transparent. The results provided by the leaky-box model are valid when used judiciously.

\end{abstract}


\section{Introduction}
Cosmic rays are generated in sources interspersed in the Galaxy and during their traversal through interstellar space their spectra and chemical composition are modified through interactions. With an understanding of these interactions and comparing the theoretical expectations of the spectra and composition with the experimental observations, we can decipher various aspects of the propagation history. The early efforts in this regard started with understanding observed abundances of elements like Li, Be, and B (in comparison with those in the universal abundance of the elements), which were several orders of magnitude greater in cosmic rays relative to elements like C, N, and O. Accordingly, they were interpreted as spallation products of the more massive elements during their traversal through interstellar matter. The early attempts to quantify this traversal were by M.V.K. Appa Rao and M.F. Kaplon (\cite{Kaplon_1963}) and by V.K. Balasubrahmanyan, E. Boldt, and R.A.R. Palmeira (\cite{Boldt_1965}). They assumed that the amount of matter traversed by cosmic rays at any given rigidity was unique, with negligible spread, similar to that in a slab of matter. It was soon realized that a calculation carried out including the loss of energy due to ionization could fit either the (Li + Be + B)/(C + N + O) ratio or the spectrum of C, N, O, but not both with the same choice of the slab thickness. The thickness needed to generate adequate Li, Be, and B caused the spectra of C, N, and O to be excessively flattened out at low energies because of the loss of energy resulting from ionization increases steeply at low energies for the nonrelativistic particles.

In order to overcome this difficulty the leaky-box model was invented in 1965 which assumed that for any rigidity there exists a broad distribution of path-lengths, called the vacuum path-length distribution, which characterizes the propagation.   Each value for the path-length leads to a different expectation for the observed spectrum. An integral over all such expectations weighted by the path-length distribution yields the prediction for the observed spectrum, which is appropriate for comparison with the observations. Based on general considerations, an exponential distribution of path lengths was assumed and two central problems were successfully analyzed to illustrate the idea: The first was a discussion of the effects of the then newly discovered universal microwave background at $\sim 3$ K on the spectrum of cosmic ray electrons (\cite{Cowsik_1966}) and the second was the spectrum and abundance ratio of Li, Be, B and C, N, O in cosmic rays (\cite{Cowsik_1967}).

This led to extensive use of the leaky-box idea (\cite{Shapiro_1970, Meneguzzi_1971,  Cesarsky_1985, Swordy_1990, Leske_1993, Jokipii_2000, Yanasak_2001, Ave_2009, George_2009, Lave_2012, Gabici_2023}) to interpret cosmic-ray data that became progressively more detailed and accurate. Instruments on platforms in space such as ACE-CRIS, Voyager, PAMELA, CRN, AMS-01, and AMS-02 on the International Space Station, ATIC, CALET and DEMPE etc. lead to rapid collection of high-quality data that needed to be interpreted.

In 1973 and 1975 Cowsik and Wilson (\cite{Cowsik_1973, Cowsik_1975}) introduced matrix methods to analyze the data on a sequence of nuclei simultaneously and also to address possible modifications to the cosmic-ray composition, subsequent to their acceleration, in cocoons surrounding the sources. This latter application required the operation by two matrices in tandem, one for the transformations in the cocoons and the other in the interstellar medium of the Galaxy from which cosmic rays escape into intergalactic space.

The use of diffusion equations to address propagation of cosmic-rays in the Galaxy was relatively rare. However, diffusion formalism became essential to address the effects of spatial and temporal discreteness of the sources of cosmic ray electrons (\cite{Shen_1971, Cowsik_1979, Nishimura_1981}). However, it was not until versatile computational packages such as GALPROP by Strong and Moskalenko (\cite{Strong_1998}) and DRAGON by Evoli et al. (\cite{Evoli_2009}) that more extensive analysis using diffusion equations came into vogue.

The main aim of this paper is to show that there is a close correspondence between these two approaches in the study of cosmic-ray propagation. The leaky-box approach is equivalent to using the impulse response function to capture the essential features of the solution to the diffusion equation. Both of these approaches require that we treat the Galaxy to be in a quasi-steady state over the time scales of propagation and that the response is linear in that a small increase in the source strength leads to a proportional increase in the observed density of cosmic rays.

This introduction is followed in Section 2.1 by setting up the diffusion equation and solving it in a model of the Galaxy with appropriate boundary conditions. Section 2.2 provides details of the source function, which is assumed to be similar to the rate of supernova explosions in various regions of the Galaxy. This is followed by an analysis in Section 2.3 that actually defines the impulse response function for the diffusion of cosmic rays in the Galaxy, as well as for a toy-model and for the leaky-box model. Section 3 is for detailing the results, discussions, and comments. Finally, Section 4 very briefly summarizes the conclusions.

\section{The correspondence between diffusion models and path length distribution}
\subsection{The Diffusion Equation and its Green's Function}
In order to investigate the correspondence, we begin with the standard diffusion equation:

\begin{equation}
      \mathbf{\nabla} \mathbf{\cdot} (\kappa\mathbf{\nabla} \mathbf{\cdot} \rho) =\frac{\partial\rho}{\partial t}
\end{equation}
where $\rho$ is the density of cosmic rays at $(\mathbf{r}, t)$ and $\kappa$ is the diffusion coefficient. We solve this in a volume bounded by two planes at $z=0$ and $z=L$ and impose the boundary condition that $\rho$ vanishes at the boundaries:
\begin{equation}
    \rho(\mathbf{r}, t) = 0\ \mathrm{at}\ z=0\ \mathrm{and}\ z=L.
\end{equation}
The solution of Equation 1. proceeds in the standard manner with the separation of variables and obtaining the eigenfunctions. In order to match the boundary conditions we choose the $z$-dependent eigenfunction normalized to unity to be 
\begin{equation}
    Z(z) = \sqrt{\frac{2}{L}}\sin\left(\frac{n \pi z}{L}\right),
\end{equation}
and obtain the Green's function for the initial condition 
\begin{equation}
    \rho(x, y, z, t = 0) = \delta(x-x_o)\delta(y-y_o)\delta(z-z_o).
\end{equation}
With these manipulations we obtain the Green's function to be
\begin{equation}
    G(x, x_o, y, y_o, z, z_o,t) = \left[\sum_{n=1}^{\infty}\frac{2}{L} \sin\left(\frac{n \pi z}{L}\right) \sin\left(\frac{n \pi z_0}{L}\right) e^{\frac{-\kappa n^2 \pi^2 t}{L^2}}\right] \times\frac{1}{4 \pi \kappa t}e^{-[(x-x_o)^2+(y-y_o)^2]/4\kappa t}.
\end{equation}
This Green's function is symmetric under the interchange of the source coordinates with no subscript and the observer's coordinates with $o$ as the subscript.

\subsection{The Distribution of Cosmic-Ray Sources}
We illustrate the correspondence with the assumption that supernovae are the sources of cosmic rays. These explosions occur close to the Galactic Plane, which corresponds to $z=L/2$ in the coordinate system set up above. To be specific, we choose the rate of supernova occurrences given by K. Ferrière (\cite{Ferriere_2001}) for both Type-I and Type-II supernovae. The expression given by Ferrière is in a cylindrical polar system of coordinates with the origin located at the Galactic Center and factors into $R$ and $\mathfrak{z}$ dependent terms, $\Gamma(R)$ and $\Gamma(\mathfrak{z})$ for each of the types of supernovae. These are expressed as variations with respect to their value at the solar system $(R_{\odot}, \mathfrak{z}_{\odot})$, namely $\Gamma_{I,_\odot}$ and $\Gamma_{II,\odot}$. Based on their prescription we adopt the following expressions:
\begin{equation}
    \Gamma_I(R, \mathfrak{z}) = \Gamma_{I, \odot}
    \left[
    e^{-\left(\frac{R-R_{\odot}}{4.5\ \mathrm{kpc}}\right)}
    e^{-\left(\frac{|\mathfrak{z}-\mathfrak{z_{\odot}|}}{0.325\ \mathrm{kpc}}\right)}
    \right]
\end{equation}
\begin{equation}
	\begin{split}
    \Gamma_{II}(R, \mathfrak{z}) =&\Gamma_{II, \odot}
    \left[\
    0.79e^{-\left(\frac{\mathfrak{z}-\mathfrak{z_{\odot}}}{0.212\ \mathrm{kpc}}\right)^2} +
    0.21e^{-\left(\frac{\mathfrak{z}-\mathfrak{z_{\odot}}}{0.636\ \mathrm{kpc}}\right)^2}
    \right] \\ \times &
    \begin{cases}
    3.55e^{-\left(\frac{R-3.7}{6.8\ \mathrm{kpc}}\right)^2}\ &\mathrm{for}\ R \leq 3.7\ \mathrm{kpc} \\ 
    e^{-\frac{R^2-R_{\odot}^2}{6.8\ \mathrm{kpc}}}\ &\mathrm{for}\ R \leq 3.7\ \mathrm{kpc},
    \end{cases}
	\end{split}
\end{equation}
or more compactly
\begin{equation}
    \Gamma_{SN}(R, \mathfrak{z}) = 
    \Gamma_{SN,\odot}\{\alpha_I\Gamma_I(R,\mathfrak{z}) + 
    \alpha_{II}\Gamma_{II}(R,\mathfrak{z})
    \}
\end{equation}
where, $\Gamma_{SN, \odot} = \Gamma_{I, \odot} + \Gamma_{II, \odot}$, $\alpha_I = \Gamma_{I, \odot} / \Gamma_{SN, \odot}$ and $\alpha_{II} = \Gamma_{II, \odot} / \Gamma_{SN, \odot}$, with $\Gamma_{I, \odot} = 7.3$ SN kpc$^{-3}$ Myr$^{-1}$ and $\Gamma_{II, \odot} = 50$ SN kpc$^{-3}$ Myr$^{-1}$. In addition, the Ferrière source function conveniently factors as $\Gamma_{j} (R, \mathfrak{z}) = \Gamma_{j}(R) \cdot \Gamma_{j}(\mathfrak{z})$ for $j=I, II$.

Keeping in mind that our prime focus is on the cosmic-ray spectral densities in the Solar neighborhood we transform this source-distribution to coordinates centered at a distance $R_{\odot}$ from the Galactic Center:
\begin{equation}
    R = \left(R_{\odot}^2 + r^2 + 2 R_{\odot}r\cos(\phi)\right)^{1/2}
\end{equation}
with obvious notation. These transformations are written as follows: $\Gamma_{SN}[R(r, \phi), \mathfrak{z}]$, $\Gamma_{I}[R(r, \phi), \mathfrak{z}]$, and $\Gamma_{II}[R(r, \phi), \mathfrak{z}]$. We note that the $\mathfrak{z}$ used by Ferrière is shifted by $L/2$ with respect to the z-coordinate used by us to derive the Green's function
\begin{equation}
    \mathfrak{z} = z - L/2,\ z_{\odot} = L/2
\end{equation}

\subsection{The Analysis}
We first consider an imaginary source of unit strength distributed across the Galaxy as described by the Ferrière function, rewritten as $\alpha_I \Gamma_I[R(r, \phi), z] + \alpha_{II} \Gamma_{II}[R(r, \phi), z]$ that is essentially the expression in Equation 9., but without the normalization of the total rate $\Gamma_{SN,\odot}$. In response to such an input operating for a short time interval $dt$ at $t = 0$ the cosmic-ray density at $R_\odot, z_0$ will be represented by
\begin{equation}
    \begin{split}
        H(R_{\odot}, z_0, t) = \sum_{i = I, II} \sum_{n=1}^{\infty}&\alpha_{i}\left[2\int_0^{\pi}\int_0^{\infty}\Gamma_{i}[R(r,\phi)]\times\frac{1}{4 \pi \kappa t}e^{-[r^2/4\kappa t]} r dr d\phi\right]  \\ & 
        \times\left[\int_0^L \Gamma_{i}(z)\sqrt{\frac{2}{L}} \sin\left(\frac{n \pi z}{L}\right)dz\right] \\ &\times\left[\sqrt{\frac{2}{L}} \sin\left(\frac{n \pi z_0}{L}\right) e^{\frac{-\kappa n^2 \pi^2 t}{L^2}}\right]
    \end{split}  
\end{equation}
We call $H$ the impulse response function at $R_{\odot}, z_o$ of the Galaxy in which supernovae are distributed according to the prescription given by Ferrière. This is displayed in Figure 1. for the choice $z_o=z_\odot=L/2$, corresponding to the assumed location of the median plane of the Galaxy in our calculations.

Before we discuss various aspects of this response function, $H$, we calculate the response functions of two other models: First, we consider a toy-model with the source function of the form

\begin{equation}
    \Gamma_{T}(r,z)=\Gamma_{T\odot}e^{-| z-\frac{L}{2}|/\lambda}
\end{equation}
Here we choose $\Gamma_{T,\odot} = [\sigma_I (R_\odot ) + \sigma_{II} (R_\odot )] $ .
This source function is independent of $r$ and the integration over $z$ yields the same number density of sources as the Ferrière function at the location of the Solar system. The toy-model has the impulse response function;
\begin{equation}
    \begin{split}
        H_T(R_{\odot}, z_0, t) = & \sum_{n=1}^{\infty}\left[\int_0^{\infty}\frac{e^{-r^2/4\kappa t}}{4 \pi \kappa t}~ 2 \pi r dr \right]  \\ &
        \times\left[\int_0^L \Gamma_{T\odot}e^{-| z-\frac{L}{2}|/\lambda}\sqrt{\frac{2}{L}} \sin\left(\frac{n \pi z}{L}\right)dz\right] \\ & 
        \times\left[\sqrt{\frac{2}{L}} \sin\left(\frac{n \pi z_0}{L}\right) e^{\frac{-\kappa n^2 \pi^2 t}{L^2}}\right].
    \end{split}  
\end{equation}
Note here the integral over $r$ yields unity, and the $z$ part of the integral yields
\begin{equation}
    \Gamma_{T\odot} \sqrt{\frac{2}{L}}\left(\frac{L^2 \lambda^2}{L^2+n^2\pi^2 \lambda^2}\right) \cdot \frac{2}{\lambda}
\end{equation}
Thus the toy-model has the impulse response function
\begin{equation}
    H_T(R_{\odot}, z_0, t) = \Gamma_{T\odot} \sum_{n=1}^{\infty}\sqrt{\frac{2}{L}}
    \left(\frac{L^2 \lambda^2}{L^2+n^2\pi^2 \lambda^2}\right) \cdot \frac{2}{\lambda} \cdot
    \sqrt{\frac{2}{L}}\sin\left(\frac{n \pi z_o}{L}\right)e^{-\frac{\pi^2 n^2 \kappa t}{L^2}}.
\end{equation}
This is also displayed in Figure 1. for $z_o = L/2$, i.e., the Galactic Plane, again without $\Gamma_{T, \odot}$, so that it represents a source of unit strength.

Finally, the leaky-box model for a unit source has the impulse response function
\begin{equation}
    H_{LB} = e^{-t/\tau}.
\end{equation}
We show this also in Figure 1. for the choice $\tau = \left(\frac{\pi^2 \kappa}{L^2}\right)^{-1} = \frac{L^2}{\pi^2 \kappa}$.

\begin{figure}
    \centering
    \includegraphics[width=0.9\linewidth]{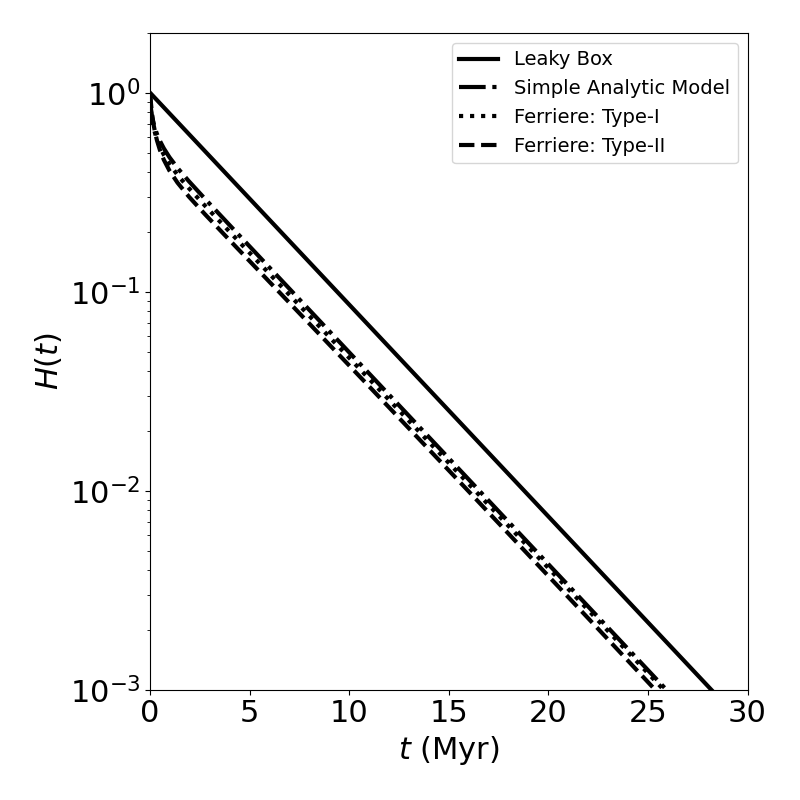}
    \caption{The impulse response functions are displayed for three different source distributions pertaining to the Ferrière distributions of Type I and Type II supernovae and a toy model with a constant radial density and an exponential decrease transverse to the Galactic plane. These show an initial steep decrease while diffusion rapidly smooths out local effects and the gradients in the source functions, followed by an exponential decay due to leakage with a time-constant $\tau = L^2 /\pi ^2 \kappa$ . A simple exponential  adopted in the leaky-box models with the with the same $\tau$  is shown for comparison.}
\end{figure}

\section{{Discussion of the analysis and results}}

The impulse response functions for the Ferrière source distribution, the toy-model, and the leaky-box model are displayed in Figure 1. The impulse response function is the temporal part of the Green's function obtained by integrating over the spatial part of the source function and, as such, adequately quantifies the responses which in our case is the cosmic-ray density near the Solar System. This is equivalent to the vacuum path-length distribution described by Cowsik et al. (\cite{Cowsik_1967}) to which the effects of spallation, energy losses, etc. could be added and a variety of cosmic ray phenomena could be investigated (\cite{Cowsik_1966, Cowsik_1967, Meneguzzi_1971, Swordy_1990, Leske_1993, Jokipii_2000, Yanasak_2001, Ave_2009, George_2009, Lave_2012}). Such systems in tandem can also be treated in a similar fashion, in what has become known in the literature as the nested-leaky-box model to address a variety of cosmic-ray phenomena (\cite{Cowsik_1973, Cowsik_1975, Cowsik_2010, Cowsik_2014, Cowsik_2016}).

Such tandem systems have been used to address the effects of cosmic ray interactions, subsequent to acceleration, in a cocoon of matter surrounding the sources. As matrices are not commutative, the order in which we apply them to analyze the data is important. A recent paper illustrates this point (\cite{Cowsik_2024}). Furthermore, this paper provides a method of deriving the source spectra and lifetimes of cosmic-rays by minimizing the abundance of the rare boron nuclei in the cosmic-ray sources, without any other assumptions, directly from the observational data.

Turning now specifically to the results obtained in the present analysis, we note the following points:

\begin{enumerate}
    \item The impulse responses for the Ferrière source functions for Type I and Type II supernovae and the simple analytic models of the distribution of sources are very similar to one another: After the initial transient responses rapidly die down, the impulse response functions settle down to a simple exponential decrease characterized by a time constant $\tau = L^2 / \pi^2 \kappa$. The normalizations are also very close. These being response functions for unit impulses, their integrals are all unity, and they correspond to the vacuum path-length distributions described by Cowsik et al.(1967).
    \item The transients are attributable to the to the contributions from the source distribution very close to the observation point and their spatial frequency content transverse to the Galactic plane. Near $t \gtrsim 0$ , the cosmic-ray densities essentially follow the source distribution. The diffusion rapidly smoothens out all but the fundamental spatial frequencies.
    \item The leaky-box model represents a steady state and has a simple exponential fall-off with the same time constant $\tau = L^2 / \pi^2 \kappa \approx 4.1$ Myr (for $\kappa = 3\times10^{28}$ cm$^2$ s$^{-1}$ and $L = 2$ kpc).
    \item The normalization of the exponential decay times for the Type I , Type II distributions (Equation 9.) both match that for the toy model, which assumes constancy in the radial direction and an exponential decrease on either side of the Galactic plane; The leaky-box model is larger by about 40\%. 
    \item Also the radial distribution of the source function plays a minor role: Near the Solar system the radial gradients are relatively small. To the extent that the radial variation can be characterized as a linear decrease with $R(r, \phi )$, it has negligible  effect on the integral of the distribution over $r$.
    \item The energy density of cosmic rays near the Solar System is estimated by multiplying the response function by $\Gamma_{SN, \odot} = (50+7.3)$ SN kpc$^{-3}$ Myr$^{-1}$, as given by Ferrière, and integrating the response function over time from $0$ to $\infty$ in order to get the steady state value. Assuming that each supernova injects $10^{50}$ ergs in cosmic rays, the integration yields:
    \begin{equation}
        \begin{split}
            \rho_{CR, \odot} &\approx 57.3 \times 10^{50}\times \tau \\
            &\approx 2.35\times10^{52}\ \mathrm{ergs\ kpc}^{-3} \\
            & \approx8\times10^{-13}\ \mathrm{ergs\ cm}^{-3}.
        \end{split}
    \end{equation}
    For a spectrum of cosmic ray proton intensities of the form $A(0.938 +E)^{-2.7}$, that is a power law in total energy, with $E$ being the kinetic energy expressed in GeV, the theoretical estimate $8\times10^{-13}\ \mathrm{ergs\ cm}^{-3}$ yields a value 
    \begin{equation}
        A = 1.356\ \mathrm{cm}^{-2}\ \mathrm{s}^{-1}\ \mathrm{sr}^{-1}\ \mathrm{GeV}^{-1}.
    \end{equation}
    This value of $A$ predicts the correct value of the observed intensities of protons at 1 GeV kinetic energy.
    \item Once the vacuum path-length distribution or the impulse response function is known, the effects of spallation, radioactive decay, and energy loss can be subsequently added on, as pointed out even in the early papers (\cite{Cowsik_1966, Cowsik_1967}).
    \item The leaky-box model describes the density of cosmic rays as a steady state average, and does not address spatial or temporal fluctuations by the discrete and transient nature of the sources. It is important to take into account spatial and temporal discreteness especially for addressing the spectral intensities of some radioactive  nuclei and the highest energy electrons in cosmic rays. This can be accomplished explicitly summing over the set of sources using the diffusion kernel and/or by a judicious combination of the diffusion approach with the leaky-box models.
\end{enumerate}

\section{Conclusion}

The leaky-box model bears close correspondence with the diffusion model approach for analyzing the cosmic-ray data, and provides a clear intuitive connection between the source function and the observed aspects of the cosmic-ray data such as the spectra and abundances of various secondaries. With the matrix formalism (\cite{Cowsik_1973, Cowsik_1975}) for both single and concatenated leaky-boxes the formalism is compact and powerful; see for example (\cite{Cowsik_2016} and \cite{Cowsik_2024} and references therein). One has to exercise great caution in the choice of the path-length distribution as some of them yield inconsistent results, even though they appear to be correct in a local context. Leaky-box models have been very useful over the decades in providing both quantitative and intuitive understanding of the cosmic-ray phenomena and are expected to serve similarly in the future.

\bibliography{bibliography}

\end{document}